\documentclass{article} 
\usepackage{nips12submit_e,times}
\usepackage{verbatim}
\usepackage{amsmath}                                                                                                                                                      
\usepackage{graphicx}                                                                                                                                                     
\usepackage{amsthm}                                                                                                                                                       
\usepackage{amssymb}                                                                                                                                                      
\usepackage{epsfig}                                                                                                                                                       
\usepackage{caption}
\usepackage{subcaption}
\usepackage{color}                                                                                                                                                        
\usepackage{url}                                                                                                                                                          

 \newcommand {\R}{\mathbb {R}}

                                                                                                                                                                     \newcommand {\B}{\mathbb {B}}

\definecolor{bbl}{rgb}{.06,.18,.91}
\definecolor{ggr}{rgb}{.06,.95,.05}
\definecolor{rrr}{rgb}{.95,.05,.06}
\definecolor{lrr}{rgb}{.95,.75,.66}
\definecolor{lbb}{rgb}{.75,.75,.96}
\definecolor{lgg}{rgb}{.75,.95,.76}
\definecolor{bwn}{rgb}{0.85,.66,.5}
\definecolor{gld}{rgb}{0.99,.99,.3}
\definecolor{grey}{rgb}{0.8,.8,.8}

\nipsfinalcopy

\title{Tree structured sparse coding on cubes}
\author{Arthur Szlam\\ 
  City College of New York\\
  \texttt{aszlam@ccny.cuny.edu}
  }
\begin{document}

\maketitle
Several recent works have discussed tree structured sparse coding \cite{jenattonProxHierarchical2010,kim_xing_group_lasso,Jacob:group,model_based_cs_arxiv}, where $N$ data points in $\R^d$ written as the $d\times N$ matrix $X$ are approximately decomposed into the product of matrices $WZ$. Here $W$ is a $d\times K$ dictionary matrix, and $Z$ is a $K\times N$ matrix of coefficients.
In tree structured sparse coding, the rows  of $Z$ correspond to nodes on a tree, and the columns of $Z$ are encouraged to be nonzero on only a few branches of the tree; or alternatively, the columns are constrained to lie on at most a specified number of branches of the tree.  

When viewed from a geometric perspective, this kind of decomposition is a ``wavelet analysis'' of the data points in $X$ \cite{Jones90,DS91,Lerman03,geometric_wavelets}. 
As each row in  $Z$ is associated to a column of $W$, the columns of $W$ also take a tree structure.   The decomposition corresponds to a multiscale clustering of the data, where the scale of the clustering is given by the depth in the tree, and cluster membership corresponds to activation of a row in $Z$.  The root node rows of $Z$ corresponds to the whole data set, and the root node columns of  $W$ are a best fit linear representation of $X$.   The set of rows of $Z$ corresponding to each node specify a cluster- a data point $x$ is in that cluster if it has active responses in those rows.  The set of columns of $W$ corresponding to a node specify a linear correction to the best fit subspace defined by the nodes ancestors; the correction is valid on the corresponding cluster.

Here we discuss the analagous construction on the binary cube $\{-1,1\}^d$.  Linear best fit is replaced by best fit subcubes. 

\section{The construction on the cube}
\subsection{Setup}
We are given $N$ data points in $\B^d=\{-1,1\}^d$ written as the $d\times N$ binary matrix $X$.
Our goal is to decompose $X$ as a tree of subcubes and ``subcube corrections''.  
A $q$ dimensional subcube $C=C_{c,I^r}$ of $\B^d$ is determined by a point $c\in \B^d$, along with a set of $d-q$ restricted indices $I^r=r_1,...,r_{d-q}$.    The cube $C_{c,I^r}$ consists of the points $b\in\B^d$ such that $b_{r_i}=c_{r_i}$ for all $r_i\in I^r$, that is 
\[C_{c,I^r}=\{b\in \B^d \text{ s.t. } b_{r_i}=c_{r_i} \,\, \forall r_i\in I^r\}.\]
  The unrestricted indices $I^u=\{1,...,d\}\setminus I^r$
  can take on either value.  
\subsection{The construction}
Here I will describe a simple version of the construction where each node in the tree corresponds to a subcube of the same dimension $q$, and a hard binary clustering is used at each stage.  Suppose our tree has depth $l$.  Then the construction consists of 
\begin{enumerate}
\item  A tree structured clustering of $X$ into sets $X_{ij}$ at depth (scale) $i\in\{1,...,l\}$ such that 
\[\bigcup_j X_{ij}=X,\]
\item and cluster representatives (that is $d-iq$-dimensional subcubes)
\[C_{c_{ij},I^r_{ij}}\]
such that the restricted sets have the property that if $ij$ is an ancestor of $i'j'$,
\[I^r_{ij}\cap I^r_{i'j'}=I^r_{ij},\]   
and \[c_{ij}(s)=c_{i'j'}(s)\] 
for all $s\in I^r_{ij}$  
 \end{enumerate}
Here each $c_{ij}$ is a vector in $\B^d$; the complete set of $c_{ij}$ roughly corresponds to $W$ from before.   However, note that each $c_{ij}$ has precisely $d-iq$ entries that actually matter; and moreover because of the nested equalities, the leaf nodes carry all the information on the branch.  This is not to say that the tree structure is not important or not used- it is, as the leaf nodes have to share coordinates.  However once the full construction is specified, the leaf representatives are all that is necessary to code a data point.
\subsection{Algorithms}
We can build the partitions and representatives starting from the root and descending down the tree as follows:  first, find the best fit $d-q$ dimensional subcube for the whole data set.  This is given by a coordinate-wise mode; the free coordinates are the ones with the largest average discrepancy from their modes.  Remove the $q$ fixed coordinates from consideration.  Cluster the reduced ($d-q$ dimensional) data using $K$ means with $K=2$; on each cluster find the best fit $(d-q)-q$ cube.  Continue to the leaves.  
\subsubsection{Refinement}
The terms  $C_{c_{ij},I^r_{ij}}$ and $X_{ij}$ can be updated with a Lloyd type alternation.  With all of the $X_{ij}$ fixed, loop through each $C$ from the root of the tree finding the best subcubes at each scale for the current partition.  Now update the partition so that each $x$ is sent to its best fit leaf cube.
\subsubsection{Adaptive $q$, $l$, etc.}
In \cite{geometric_wavelets}, one of the important points is that many of the model parameters, including the $q$, $l$, and the number of clusters could be determined in a principled way.  While it is possible that some of their analysis may carry over to this setting, it is not yet done.  However, instead of fixing $q$, we can fix a percentage of the energy to be kept at each level, and choose the number of free coordinates accordingly.

\section{Experiments} 
We use a binarized the MNIST training data by thresholding to obtain $X$.  Here $d=28^2$ and $N=60000$.  
Replace $70\%$  of the entries in $X$ with noise sampled uniformly from $\{-1,1\}$, and train a tree structured cube dictionary with $q=80$ and depth $l=9$.   
The subdivision scheme used to generate the multiscale clustering is $2$-means initialized via randomized farthest insertion \cite{Arthur:2007:KAC:1283383.1283494}; this means we can cycle spin over the dictionaries \cite{coifmandonoho95}, to get many different reconstructions to average over.  In this experiment the reconstruction was preformed 50 times for the noise realization.  The results are visualized below.

\begin{figure}
	\centering
	\begin{minipage}{.5\linewidth}
		\includegraphics[width=.95\textwidth]{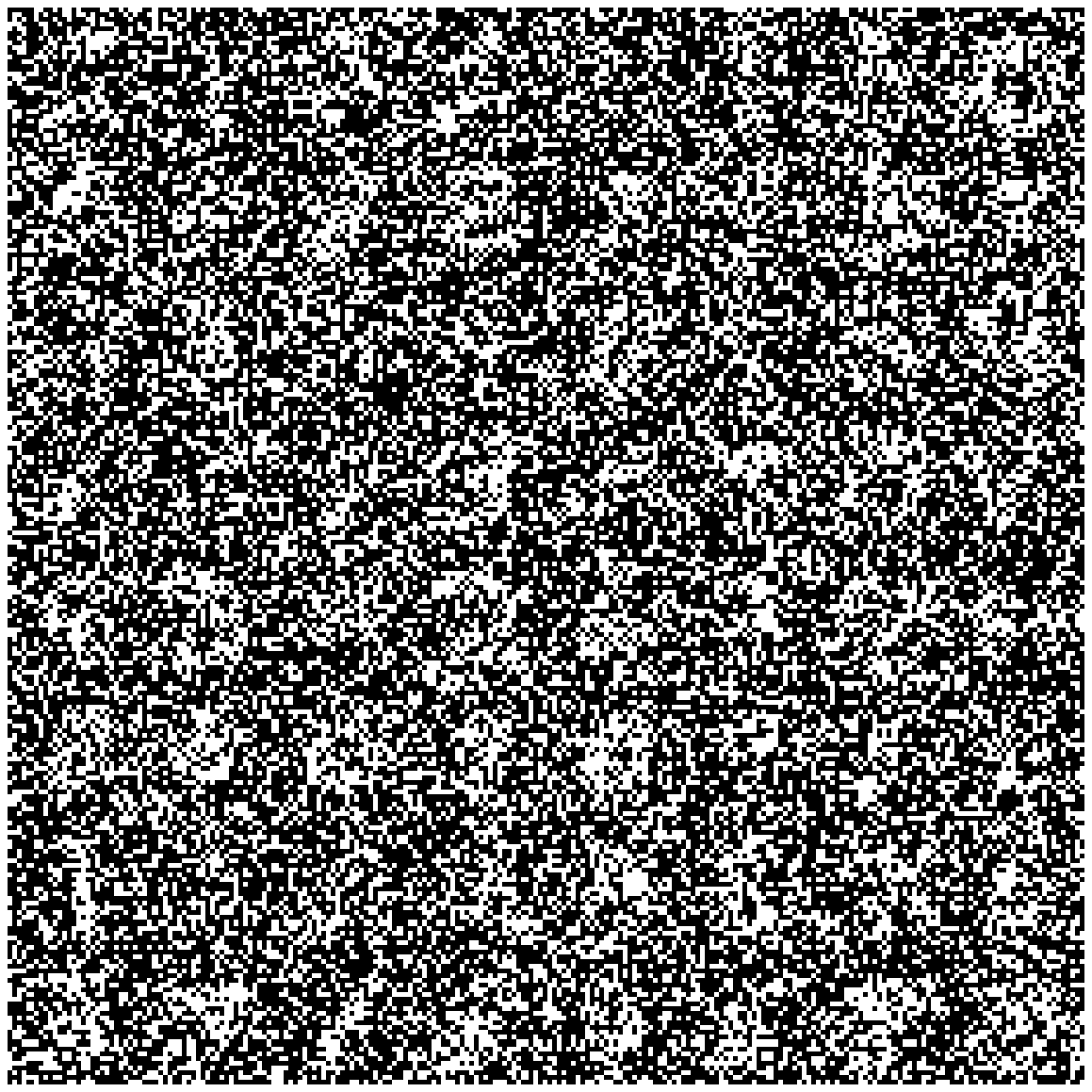}
	\end{minipage}%
	\begin{minipage}{.5\textwidth}
		\includegraphics[width=.95\textwidth]{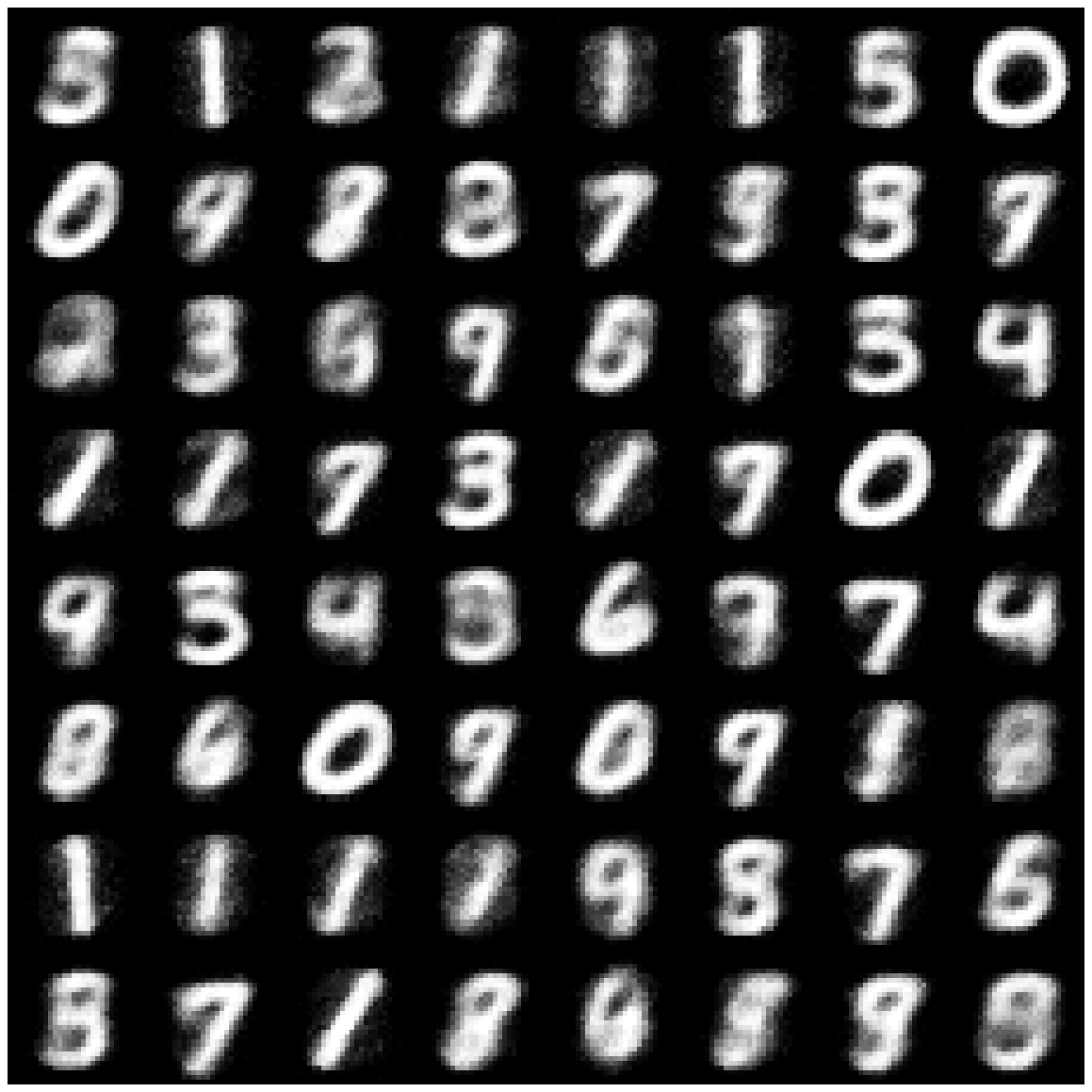}
	\end{minipage}%
	\newline
     \begin{minipage}{.5\textwidth}
		\includegraphics[width=.95\textwidth]{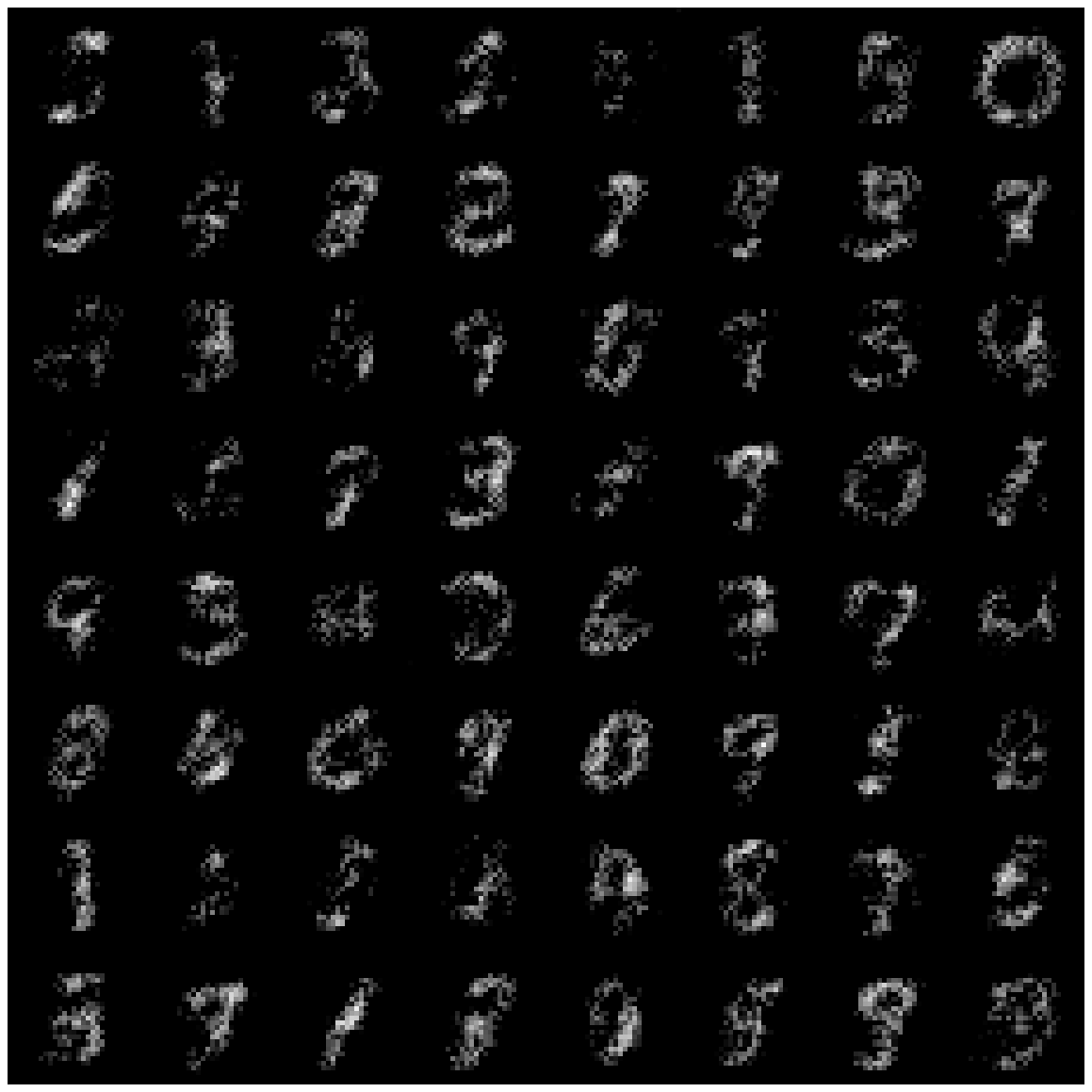}
	\end{minipage}%
	\begin{minipage}{.5\textwidth}
		\includegraphics[width=.95\textwidth]{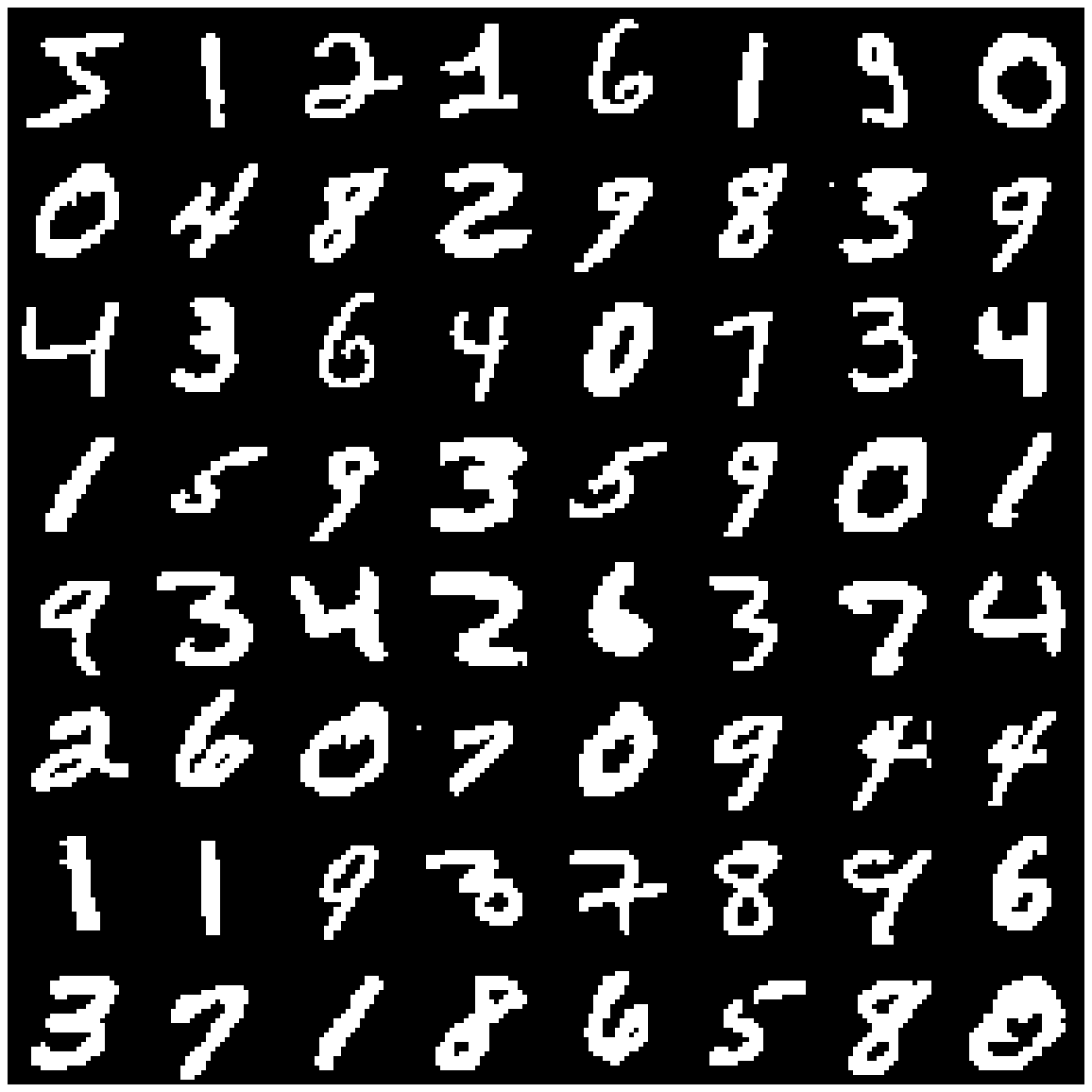}
	\end{minipage}%
		\caption{Results of denoising using the tree structured coding.  The top left image is the first 64 binarized MNIST digits after replacing $70\%$ of the data matrix with uniform noise.  The top right image is recovered, using a binary tree of depth $l=9$ and $q=90$, and 100 cycle spins, thus the non-binary output, as the final result is the average of the random clustering initialization (of course with the same noise realization).  The bottom left image is recovered using robust pca \cite{DBLP:journals/jacm/CandesLMW11}, for comparison.  The bottom right is the true binary data.}
		\label{fig:mnist_reconstruct}
\end{figure}

\bibliographystyle{plain}
\bibliography{article}
\end{document}